\documentclass[aps,twocolumn,superscriptaddress,letterpaper]{revtex4}
\usepackage{graphicx,amssymb,bm,natbib,amsfonts,amsmath}
\usepackage{hyperref}

\begin{document}

\title{Probing the band structure of LaTe$_2$ using angle resolved photoemission spectroscopy}

\author{D.R. Garcia}
\affiliation{Department of Physics, University of California,
Berkeley, CA 94720, USA}
\affiliation{Materials Sciences Division,
Lawrence Berkeley National Laboratory, Berkeley, CA 94720, USA}
\author{S.Y. Zhou}
\affiliation{Department of Physics, University of California, Berkeley, CA 94720, USA}
\affiliation{Materials Sciences Division, Lawrence Berkeley National Laboratory, Berkeley, CA 94720, USA}
\author{G.-H. Gweon$^*$}
\affiliation{Department of Physics, University of California,
Berkeley, CA 94720, USA}
\author{M.H. Jung}
\affiliation{National Research Laboratory for Material Science, KBSI, 52 Yeoeun-Dong Yusung-Gu, Daejeon 305-333, Korea}
\author{Y.S. Kwon}
\affiliation{Department of Physics, Sung Kyun Kwan University, Suwon 440-746, South Korea}
\author{A. Lanzara}
\affiliation{Department of Physics, University of California,
Berkeley, CA 94720, USA}
\affiliation{Materials Sciences Division, Lawrence Berkeley National Laboratory, Berkeley, CA 94720, USA}

\date{\today}

\begin{abstract}
With the current interest in the rare-earth tellurides as 'high temperature' charge density wave materials, a greater understanding of the physics of these systems is needed, particularly in the case of the ditellurides.  We report a detailed study of the band structure of LaTe$_2$ in the charge density wave state using high-resolution angle resolved photoemission spectroscopy (ARPES).  From this work we hope to provide insights into the successes and weaknesses of past theoretical study as well as helping to clear up prior ambiguities by providing a firm experimental basis for future work in the tellurides.
\end{abstract}

\maketitle
Having been the subject of research for over half a century, a charge density wave (CDW) transition relates to a balance between electronic energy and lattice structural stability.  Below a critical temperature, the system finds it energetically favorable to introduce a new periodic ordering, allowing the Fermi surface (FS) to be gapped, and lowering the overall electronic energy.  Because this ordering is mediated by strong electron-phonon interactions, such systems can provide excellent opportunities for theoretical investigation toward how strongly-correlated electron-phonon systems behave and its effects on band structure.  Furthermore, when one relates CDW and other charge ordering physics to systems exhibiting superconductivity, the interest becomes more pressing and exciting \cite{Morris,Nunezregueiro,Singh,Fang,Morosan,Jung1}.  

Recently the rare earth di- and tritelluride systems have attracted great interest due to their low dimensionality and the recent discovery of a pressure-induced superconducting state \cite{Jung1} competing with a CDW phase and with anti-ferromagnetic order.  This interplay makes the tellurides the ideal system to investigate the consequences that the competition between charge density wave, antiferromagnetism and superconductivity has on the fermionic excitation at the Fermi energy and to provide a deeper understanding of how superconductivity results from such interplay, an issue of great interest in the solid state community.  

The existence of a CDW phase in the ditellurides has been first established by transmission electron microscopy (TEM) \cite{DiMasi, Shin} and single crystal X-ray diffraction \cite{Stowe} experiments.  TEM measurements have reported a long range distortion $\bf{q}$ = .5$\bf{a}$$^*$ \cite{DiMasi} similar to the diselenides \cite{Marcon}, while single crystal X-ray diffraction expanded on this suggesting a larger $2\times2\times1$ superstructure \cite{Stowe}.  Most recently, the TEM work of Shin et al. demonstrated a four-fold symmetric superstructure with a modified $\bf{q}$=.484$\bf{a}$$^*$ and proposed an additional CDW wave vector $\bf{q}$=$.6\bf{a}$$^*$+.2$\bf{b}$$^*$\cite{Shin}.  

Although traditional scattering and tunneling techniques can reveal structural modulations due to CDW phenomena, a complete insight into the CDW phase and its formation demands a direct probe of the electronic structure.  To accomplish such a study, angle resolved photoemission spectroscopy (ARPES) is unique among experimental tools. Using soft X-ray light, valence band electrons are photoemitted and their in-crystal momentum and energy are measured. This allows one to directly probe the valence band electronic states, resolving the actual FS band structure and band dispersions at higher binding energies.
ARPES is well qualified for studying such materials as it has already proved itself in the study of SmTe$_{3}$ \cite{Gweon} and CeTe$_{3}$ \cite{Brouet}, given their quasi-2D structure.   Yet, only quite recently has ARPES been used to study rare earth ditellurides such as LaTe$_{2}$ \cite{Shin}.
CDW physics can manifest itself in ARPES data in at least three ways: 1) Observing FS band structure which can be connected or "nested" by a $\bf{q}$$_{CDW}$  2) Gapping of near E$_F$ band structure, which represents the energy gained by the instability 3) The appearance of "shadow bands" where the action of the CDW creates new states by shifting the main band structure by $\bf{q}$$_{CDW}$. 
Thus, in order to achieve any insight into CDW physics, we must first begin by understanding the near E$_F$ band structure.

\begin{figure*}
\includegraphics[width=14.0 cm]{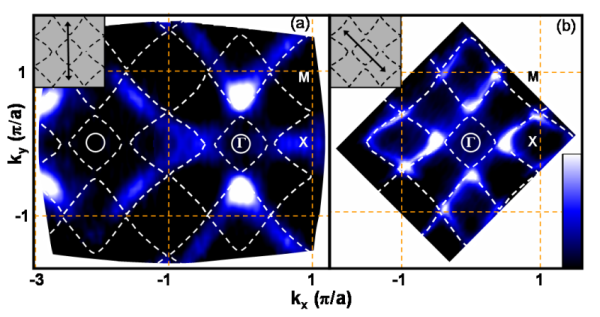}
\caption{(a) ARPES constant energy intensity map averaged in energy from E$_{F}$ to 80meV binding energy and symmetrized over x = 0.  Beam energy was 55eV with a polarization reflected in the inset.  Orange dashed lines indicate the Brillouin zone boundaries.  By comparison with the non-CDW LDA FS calculation \cite{Shim}, indicated as white dashed lines, we find moderate agreement.  However, as we approach a photon energy of 110eV, more features are revealed. (b) ARPES constant energy map, unsymmetrized, averaged from E$_{F}$ to 100meV binding energy using 110eV photons.}
\end{figure*}

    The purpose of this article is to assist the situation by providing an in-depth picture of the band structure of LaTe$_{2}$ while the specifics of CDW physics are addressed in other work \cite{Garcia}.  We make comparison to prior LDA calculation to aid theoretical work and to help better understand band structure features.  In our comparison, we also provide possible renormalizations to the current LDA work for greater consistency with our results.  We conclude with some observations about surface structure using low energy electron diffraction (LEED) which may be of importance to future ARPES or other surface sensitive work on the rare earth telluride systems.  

\begin{figure}
\includegraphics[width=9.0 cm]{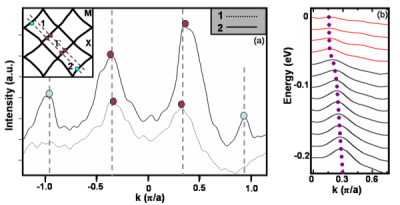}
\caption{(a) Momentum distribution curves (MDCs) taken (1) along the M-$\Gamma$-M direction and (2) parallel to M-$\Gamma$-M but slightly shifted.  From these we can discern the twin peaks of the inner diamond while the spectral of the bands near the M point is strangely suppressed but only along the high symmetry direction. (b) MDCs along $\Gamma$-M at increasing binding energies showing the dispersion of the inner diamond band up into the energy range of Figure 1.  Red MDCs indicate where the curve becomes dispersionless}
\end{figure}

\begin{figure*}
\includegraphics[width=16.0 cm]{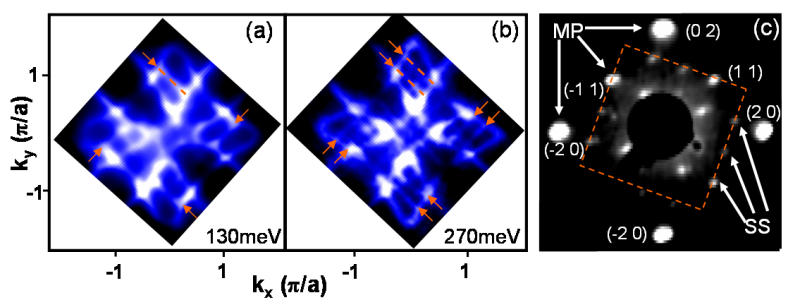}
\caption{(a) ARPES constant energy image plot centered at 130meV and integrated over 60meV.  Orange arrows indicate bands near the X points which break mirror symmetry. (b) A similar plot centered at 270meV and integrated over 80meV showing how the non-mirror symmetric bands disperse by splitting in two. (c) LEED done on LaTe$_{2}$ indicating both main peaks (MP) but also a superstructure (SS) which breaks mirror symmetry.}
\end{figure*}

    ARPES and LEED data were taken on single crystals of LaTe$_{2}$ using beam lines 7.0.1 and 10.0.1 at the Advanced Light Source of the Lawrence Berkeley National Laboratory.  These beam lines were equipped with Scienta SES100 and R4000 electron analyzers.  A total energy resolution of 40 meV or better was used, and the total angular resolution was set to 0.35 degrees.  Our samples were cleaved {\em in situ} with a base pressure better than $7\times 10^{-11}$ torr at low temperatures.  Recently the question has arisen about whether the differences in sample preparation may be responsible for differences observed in the superstructure by XRD and TEM \cite{Shin}.  The samples used for this study were grown using two different techniques: 1)Mineralization of a stoichiometric binary mixture of elements \cite{M.H.Jung2} and 2)A high-temperature Bridgeman method explained by Kwon et al. \cite{Y.S.Kwon}. Our ARPES work was unable to discern any difference originating from the two preparations.

\begin{figure}
\includegraphics[width=8.0 cm]{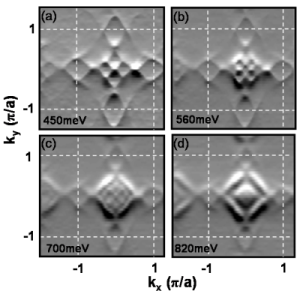}
\caption{(a)-(d) Constant energy cuts integrated over 60meV, with first derivative used to enhance band edges, centered at binding energies 450, 560, 700, and 820meV respectively.}  
\end{figure}   

\begin{figure*}
\includegraphics[width=16.0 cm]{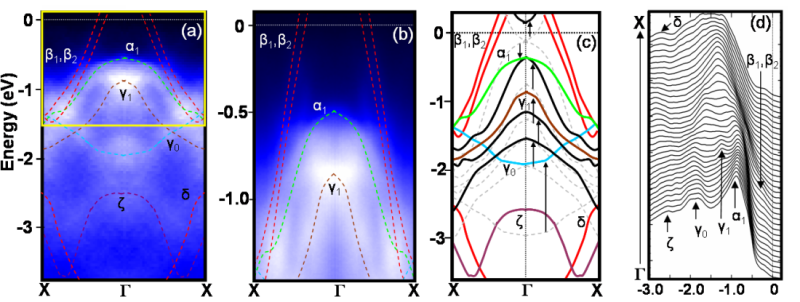}
\caption{(a) ARPES image plot showing dispersions along X-$\Gamma$-X down to nearly 4eV in binding energy. (b) ARPES image plot showing dispersions along X-$\Gamma$-X down to nearly 1.5eV in binding energy.  Overlaid on both plots is the predicted LDA band structure which, in some cases, is renormalized based on the data.  (c) LDA band structure calculation along X-$\Gamma$-X.  Bands that have been renormalized are indicated by arrows showing the proposed shift in energy from their original locations (gray dashed lines).  (d) Energy distribution curves (EDC's) down to 3eV in binding energy taken along $\Gamma$-X direction.}
\end{figure*}

     Figure 1 shows the unsymmetrized Fermi surface maps of LaTe$_2$ for two different polarization vectors and photon energies. In panel a, we show data taken at 55eV using a polarization vector along the $\Gamma$-X direction (see inset).  The map was collected up to the second Brillouin zone (BZ) and is integrated from E$_{F}$ to 80meV in binding energy.  The non-CDW LDA calculation \cite{Shim} (white dashed line) is overplotted on the same figure for comparison and shows an overall good agreement with the experimental data both in the first and higher BZs.  This supports the prediction that the near E$_F$ band structure can be thought of 1D bands arising from the 5p$_{x}$ and 5p$_{y}$ orbitals in the Te square planes, as well as confirming the tetragonal picture expected for LaTe$_2$.  A closer analysis reveals the absence of the two features centered around the $\Gamma$ point, a small electron pocket (solid white line) and an `inner diamond' band (dashed line). 

In panel b, we show data taken with a different polarization vector, along the $\Gamma$-M direction (see inset), at 110eV photon energy.  We observe that when the photon energy is tuned close to the La and Te adsorption edges (La 4d$_{3/2}\approx$105eV, Te 4p$_{3/2}\approx$103.3eV) we can resolve, although still weak, the inner diamond at the $\Gamma$ point, predicted by LDA but not observed before \cite{Shin} within this energy range.  We note, however, that despite our extensive search over a wide range of photon energies (between 80 and 200eV) and different polarization conditions, we were unable to resolve the electron pocket centered at the $\Gamma$ point.  In addition, we note that while the bands of panel b show a better overall agreement to theory, we suspect that the slight shift in band position between panels a and b is probably due to differences in the the sample surface stoichiometry over time. 

\begin{figure*}
\includegraphics[width=16.0 cm]{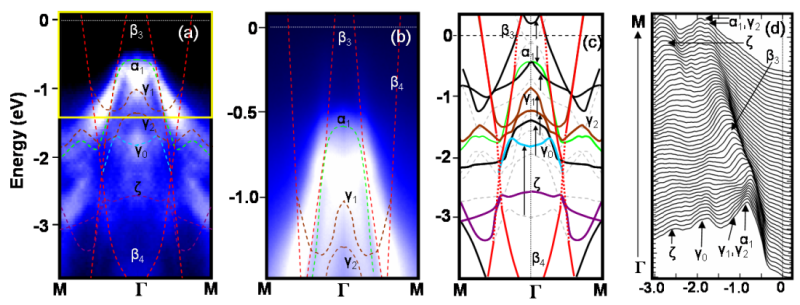}
\caption{(a) ARPES image plot showing dispersions along M-$\Gamma$-M down to nearly 4eV in binding energy. (b) ARPES image plot showing dispersions along M-$\Gamma$-M down to nearly 1.5eV in binding energy.  Overlaid on both plots is the predicted LDA band structure which, in some cases, is renormalized based on the data.  (c) LDA band structure calculation along M-$\Gamma$-M.  Bands that have been renormalized are indicated by arrows showing their proposed shift in energy from their original locations (gray dashed lines).  (d) Energy distribution curves (EDC's) down to 3eV in binding energy taken along $\Gamma$-M direction.}
\end{figure*} 

\begin{figure*}
\includegraphics[width=16.0 cm]{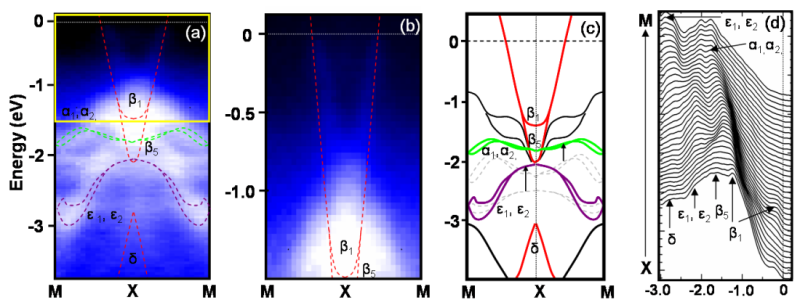}
\caption{(a) ARPES image plot showing dispersions along M-X-M down to nearly 4eV in binding energy. (b) ARPES image plot showing dispersions along M-X-M down to nearly 1.5eV in binding energy.  Overlaid on both plots is the predicted LDA band structure which, in some cases, is renormalized based on the data.  (c) LDA band structure calculation along M-X-M.  Bands that have been renormalized are indicated by arrows showing their proposed shift in energy from their original locations (gray dashed lines).  (d) Energy distribution curves (EDC's) down to 3eV in binding energy taken along X-M direction.}
\end{figure*}

The existence of the inner diamond band feature at these binding energies is further supported by momentum distribution curves (MDCs) shown in Figure 2a for cuts both along the M-$\Gamma$-M direction and slightly shifted, cuts 1 and 2 respectively.  In these cuts, the two peaks of the inner diamond band (marked by violet circles) can be well resolved.  For cut 2, we can resolve two additional peaks near $\pi$/a, corresponding to the bands near the M point (marked by light blue circles).  These are not resolved in cut 1 probably due to a matrix element effect.  In Figure 2b, we show the MDCs curves for a cut along the $\Gamma$-M direction at increasing binding energy.  The peak in the MDCs spectra, corresponding to the inner diamond band and still marked by violet dots, can be clearly distinguished in the entire energy range, from E$_F$ all the way to 200 meV binding energy.  This peak disperses from 200 meV up to 50 meV, where it suddenly stops (see peak position in the red curves), suggesting that the inner diamond is gapped by 50 meV, due to the CDW formation \cite{Garcia}.  This is contrary to prior published work, taken at photon energy far from the resonance energy, where it has been suggested that the inner diamond band are gapped by as much as 600meV in binding energy \cite{Shin}.

    In Figure 3, we show how the band structure evolves as we increase the binding energy, still remaining near the Fermi energy.  Figure 3a presents a constant energy, k-space plot of the bands around 130meV in binding energy.  A closer examination of the four X points of the Brillioun zone reveals the onset of unpredicted features, indicated by the orange arrows.  Particularly fascinating about these features is that they appear to break the mirror symmetry of the system and are completely unexpected from band structure calculation.  These features appear to be more pronounced at lower temperatures, and they are dispersive, splitting as the binding energy increases (panel 3b).  To shed more light on the possible origin of these bands, we have performed low energy electron diffraction (LEED) measurements on the sample surface, the results of which are shown in Figure 3c.  The main Bragg spots due to the orthorhombic unit cell are indicated and obey an h + k = even condition \cite{DiMasi,Shin}.   In addition to these main peaks, one can measure satellite peaks appearing to form a second superstructure which indeed breaks mirror symmetry.  This allows us to at least confirm the sample surface does break mirror symmetry.  Additional work has already suggested that this superstructure does not correspond to a CDW state of the material \cite{Garcia}.  Thus, a possible explanation for this superstructure comes from comparison with the structurally similar LaSe$_{2}$.  When the material is slightly off in stoichiometry, LaTe$_{1.9}$, the system is known to exhibit an ordered-defect superstructure which breaks mirror symmetry in the crystal's square chalcogenide planes \cite{Grupe,Lee}.  It is possible that the stoichiometry of our samples also is slightly less then perfect, either from synthesis or through change over time, and the observed band structure is due to a surface state forming on a non-mirror symmetric LaTe$_{2}$ surface. 

In Figure 4, we show the evolution of the band structure as binding energy is increased up to approximately 1eV.  In particular, we show the first derivative of the ARPES constant energy maps.  As the binding energy increases, we can clearly distinguish the onset of complex patterns centered at the $\Gamma$ point, which evolve into a checkerboard-like structure in the momentum space as the binding energy increases toward 700meV (panel 4c).  These structures maintain the four-fold symmetry seen near E$_F$.  To explain whether the onset of this complex structure is simply due to band structure or reflects some hidden order such as a CDW order, it is important to carefully examine the band structure along high symmetry direction and compare it with known LDA calculation \cite{Shim}.  
We first note that these complex patterns seem to be strongly sensitive to photon energy.  In particular they can be enhanced for photon energies close to the La 4d$_{3/2}$ adsorption edge ($\approx$105.3eV), suggesting that they are related to the LaTe layers.

To investigate the origin of these complex patterns in the momentum space, Figures 5-7 show a detailed analysis of the experimentally measured band structure along the high symmetry directions.  Figures 5a-b shows the raw ARPES image plots along the X-$\Gamma$-X direction, with the associated energy distribution curves (EDCs) shown in panel d.  The direct comparison with the LDA calculation (panel c) show that far more bands are observed experimentally than the one predicted by the non-renormalized LDA within the 400-850meV energy range.  However, bands arising from the square Te planes ($\beta$$_{1}$, $\beta$$_{2}$, and $\delta$) are, for the most part, well modeled by theory and require no renormalization.  However, bands arising from the interlayer LaTe blocks (all other bands in the figure) require renormalization.  Figure 5c illustrates our hypothesized renormalizations of the LDA band structure.  For the following reasons, we propose that the LaTe bands are being compressed into a smaller energy range, with at least three bands peaking at $\Gamma$ between 0.5 and 1 eV in binding energy.  
   First, $\alpha$$_{1}$ was expected to rise to lower binding energy and nearly touch the circular electron pocket of earlier discussion.  However, since this pocket cannot be resolved and we see no spectral weight at the $\Gamma$ point until nearly 400meV, we infer that the hybridization between the two bands was simply underestimated by theory.  A greater energy gap between the bands both forces the electron pocket above E$_{F}$ (where ARPES cannot observe it) and pushes the $\alpha$$_{1}$ to higher binding energies, around 500meV.  Secondly, it is important to be able to identify the $\gamma$$_{0}$ band in our data because it is the highest binding energy band of the LaTe block in our energy window of interest.  We find a band in the data with the appropriate curvature but it has been renormalized to a far smaller binding energy as indicated in panel c.  We take this as suggesting that the other LaTe block bands above it may require a similar renormalization pushing them all to lower binding energies and compressing them closer the energy range between 0.5 and 1.0eV.  Thus, $\alpha$$_{1}$ and $\gamma$$_{0}$ provide the `bookends' which define the new, tighter energy range which the LaTe block bands exist in.  

   Examining the other high symmetry directions, M-$\Gamma$-M in Figure 6, and M-X-M in Figure 7, we obtain a similar picture.  The M-$\Gamma$-M band structure does reasonably well at explaining the $\beta$$_{3}$ and $\beta$$_{4}$ bands, responsible for the inner diamond and near M point band structure respectively.  However, we can again identify the $\alpha$$_{1}$ and $\gamma$$_{0}$ bands to suggest the aforementioned LaTe block band renormalization.  It is also worth observing the strong increase in the spectral weight of the $\beta$$_{3}$ band (possibly due to the presence of $\alpha$$_{1}$) at around 500meV as partly responsible for the large gap associated with this inner diamond band \cite{Shin}.  Also, we suspect the $\zeta$ band may also require a renormalization to help explain the data particularly near the M point in Figure 6.  However, we do not feel that our data provide enough insight to intelligently propose one, and thus we leave the band unchanged in Figure 6.  The M-X-M experimental band structure also demonstrates a reasonable level of agreement with theory in the case of the Te plane bands but renormalization is needed for the LaTe block bands.  However, theory calculated for this symmetry direction appears far more discordant with actual experimental results than other symmetry directions.  Whether the absence of bands is due to matrix elements or a particularly non-trivial renormalization, is unclear and requires more theoretical insight.  It is our opinion that this renormalization is responsible and can provide an explanation for the complex checkerboard-like structure observed in the first derivative constant energy maps discussed in Figure 4.
       
  To summarize, we have attempted to present a clear and comprehensive picture of the experimental band structure for LaTe$_{2}$ using the ARPES technique.  We show that the band structure near E$_F$ can be well modeled by non CDW-LDA band structure calculations and is dominated by contributions from the Te(1) layers.  We found that the inner diamond structure is apparently gapped by a gap roughly an order of magnitude smaller than previously reported.  As the binding energy increases, we have identified the onset of an order that breaks mirror symmetry on the crystal surface.  We suspect this order is responsible for the formation of mirror asymmetric surface states seen in ARPES which are not believed to be part of the main band structure.  At higher binding energy, we find that contribution from other planes become important, giving rise to complex checkerboard-like patterns in the momentum space centered around $\Gamma$.   To account for these unusual patterns, we believe the LDA band structure, particularly the LaTe block bands, needs to be renormalized. We provided a possible renormalization which explains the experimental data and suggests many of these bands may be compressed into a smaller energy range than previously thought.  Of course, extended theoretical work is needed to prove this point.

This work was supported by the Director, Office of Science, Office of Basic Energy Sciences, Division of Materials Sciences and Engineering of the U.S Department of Energy under Contract No.~DEAC03-76SF00098 and by the National Science Foundation through Grant No. DMR03-49361.  The Advanced Light Source is supported by the Director, Office of Science, Office of Basic Energy Sciences of the U.S. Department of Energy under Contract No.\ DE-AC02-05CH11231. 

$^*$ Current address: Department of Physics, University of California, Santa Cruz, CA, 95060

\end{document}